\documentclass{emulateapj}
\usepackage{psfig}

\newcommand{\fig}[1]{Figure~\ref{#1}}
\newcommand{\tab}[1]{Table~\ref{#1}}

\newcommand{\sst}{SST}                   %latex for sst moniker
\newcommand{\hst}{HST}                   %latex for hst moniker
\newcommand{\galfit}{\texttt{GALFIT}}    %latex for galfit moniker
\newcommand{\mopex}{\texttt{mopex}}      %latex for mopex moniker
\newcommand{\surfsup}{SURFS UP}          %latex for surfsup moniker
\newcommand{\sersic}{S\'ersic}           %latex for Sersic
\newcommand{\pygfit}{\texttt{PyGFIT}}    %latex for PyGFIT
\newcommand{\lephare}{\texttt{Le~Phare}} %latex for lephare

\newcommand{\nframes}{277}              %number of frames in this paper
\newcommand{\nclust}{10}                 %N_cluster in SURFSUP
\newcommand{\ngal}{10}                   %number of galaxies in Hall'12

\newcommand{\psilensed}{\ensuremath{16^{+16}_{-8}}}  %linear, lensed SFR (M/yr)
\newcommand{\psiunlensedest}{\ensuremath{1.3}}    %linear, unlsned estimated
\newcommand{\massunlensedest}{\ensuremath{2.0\times10^9}} %lin,unlens,estim
\newcommand{\ssfrestpergyr}{\ensuremath{0.7}} %lin,unlens,estim
\newcommand{\aveciiflux}{\ensuremath{1.6~\mbox{mJy}}} %depends on SFR
\newcommand{\ciiflux}{\ensuremath{10^{-17}~\mbox{erg}~\mbox{s}^{-1}~\mbox{cm}^{-2}}}
\newcommand{\avefirflux}{\ensuremath{12~\mu\mbox{Jy}}}
\newcommand{\firflux}{\ensuremath{4\times10^{-15}~\mbox{erg}~\mbox{s}^{-1}~\mbox{cm}^{-2}}}

\shorttitle{Stellar Masses from \surfsup}
\shortauthors{Ryan Jr. et al.}

\begin{document}

\title{Measuring the Stellar Masses of $z\!\sim\!7$ Galaxies with Spitzer Ultrafaint Survey Program (\surfsup)\footnote{Observations were carried out using Spitzer Space Telescope, which is operated by the Jet Propulsion Laboratory, California Institute of Technology under a contract with NASA.  Also based on observations made with the NASA/ESA Hubble Space Telescope, obtained at the Space Telescope Science Institute, which is operated by the Association of Universities for Research in Astronomy, Inc., under NASA contract NAS 5-26555 and NNX08AD79G.  These observations are associated with programs Spitzer \# 3550, \#60034, \#90009, HST \#GO 10200, GO 10863, 11099, and 11591, and ESO Large Program 181.A-0485.}}

\email{rryan@stsci.edu}
\author{R. E. Ryan Jr.\altaffilmark{2},
A. H. Gonzalez\altaffilmark{3},
B. C. Lemaux\altaffilmark{4},
M. Brada\v{c}\altaffilmark{5},
S. Casertano\altaffilmark{2},
S. Allen\altaffilmark{6},
B. Cain\altaffilmark{5},
M. Gladders\altaffilmark{7},
N. Hall\altaffilmark{5},
H. Hildebradt\altaffilmark{10},
J. Hinz\altaffilmark{9},
K.-H. Huang\altaffilmark{5},
L. Lubin\altaffilmark{5},
T. Schrabback\altaffilmark{10},
M. Stiavelli\altaffilmark{2},
T. Treu\altaffilmark{11,12},
A. von der Linden\altaffilmark{6,13},
D. Zaritsky\altaffilmark{9}}
\altaffiltext{2}{Space Telescope Science Institute, 3700 San Martin Dr., Baltimore, MD 21218}
\altaffiltext{3}{Department of Astronomy, University of Florida, 211 Bryant Space Science Center, Gainesville, FL 32611}
\altaffiltext{4}{Aix Marseille Universit\'e, CNRS, LAM (Laboratoire d'Astrophysique de Marseille) UMR 7326, 13388, Marseille, France}
\altaffiltext{5}{University of California, One Shields Ave., Davis, CA 95616}
\altaffiltext{6}{Kavli Institute for Particle Astrophysics and Cosmology, Stanford University, 452 Lomita Mall, Stanford, CA 94305-4085}
\altaffiltext{7}{Department of Astronomy and Astrophysics, University of Chicago, 5640 South Ellis Avenue, Chicago, IL 60637}
\altaffiltext{8}{Department of Physics and Astronomy, University of British Columbia, 6224 Agricultural Road, Vancouver, BC V6T 1Z1, Canada}
\altaffiltext{9}{Steward Observatory, University of Arizona, 933 N.~Cherry Ave., Tucson, AZ 85721}
\altaffiltext{10}{Argelander Institute for Astronomy, University of Bonn, Auf dem H\"{u}gel 71, D-53121 Bonn, Germany}
\altaffiltext{11}{Department of Physics, University of California, Santa Barbara, CA 93106}
\altaffiltext{12}{Kavli Institute of Theoretical Physics, University of California, Santa Barbara, CA 93106-4030}
\altaffiltext{13}{Dark Cosmology Centre, Niels Bohr Institute, University of Copenhagen, Juliane Maries Vej 30, 2100 Copenhagen {\O}, Denmark}

\begin{abstract}

We present Spitzer/IRAC observations  of nine $z'$-band dropouts highly
magnified  ($2\!\lesssim\!\mu\!\lesssim\!12$) by  the  Bullet Cluster.
We combine  archival imaging with our Exploratory  program (SURFS UP),
which results  in a total  integration time of $\sim\!30$~hr  per IRAC
band. We detect ($\gtrsim\!3\sigma$)  in both IRAC bands the brightest
of  these high-redshift  galaxies,  with $[3.6]\!=\!23.80\pm0.28$~mag,
$[4.5]\!=\!23.78\pm0.25$~mag,   and   $(H-[3.6])\!=\!1.17\pm0.32$~mag.
The remaining eight galaxies are undetected to $[3.6]\!\sim\!26.4$~mag
and     $[4.5]\!\sim\!26.0$~mag     with     stellar     masses     of
$\sim\!5\times10^{7}$~M$_{\odot}$.    The  detected   galaxy   has  an
estimated magnification of $\mu\!=\!12\pm4$, which implies this galaxy
has an  ultraviolet luminosity of $L_{1500}\!\sim\!0.3\;L^*_{z\!=\!7}$
--- the lowest luminosity {\it  individual} source detected in IRAC at
$z\!\gtrsim\!7$.   By modeling the  broadband photometry,  we estimate
the    galaxy    has    an    intrinsic   star-formation    rate    of
$\mbox{SFR}\!\sim\!\psiunlensedest~\mbox{M}_{\odot}$~yr$^{-1}$      and
stellar  mass  of $M\!\sim\!\massunlensedest~\mbox{M}_{\odot}$,  which
gives     a     specific      star-formation     rate     of     ${\rm
  sSFR}\!\sim\!\ssfrestpergyr$~Gyr$^{-1}$.    If   this   galaxy   had
sustained this star-formation rate  since $z\!\sim\!20$, it could have
formed the observed stellar mass (to within a factor of $\sim\!2$), we
also  discuss   alternate  star-formation  histories   and  argue  the
exponentially-increasing  model is  unlikely.  Finally,  based  on the
intrinsic star-formation  rate, we estimate  this galaxy has  a likely
[\ion{C}{2}]       flux      of       $\langle      f_{[\mbox{C~{\tiny
        II}}]}\rangle\!=\!\aveciiflux$.
\end{abstract}


\section{Introduction} \label{sec:intro}

Determining  the details of  cosmic reionization  of hydrogen  at high
redshift  is a  central  question to  modern  cosmology. Although  the
observed  optical  depth to  Thompson  scattering \citep{hinshaw}  and
complete Gunn-Peterson troughs  \citep{bob} suggest that instantaneous
reionization  occurred  around  $z\!\sim\!10$  and  was  completed  by
$z\!\sim\!6$, the sources responsible for the ionizing radiation are
far from clear.  Although dwarf galaxies are sufficiently numerous and
energetic          to          reionize          the          Universe
\citep[e.g.][]{yan04,bou06,saw06,lem09},  it   is  uncertain  how  the
ionizing   photons   escape   such  galaxies   \citep[e.g.][]{shap06}.
Consequently  tracing  the physical  properties  of  the dwarf  galaxy
population  into the  neutral  epoch is  key  in understanding  cosmic
reionization  \citep[e.g.][]{font12}, and  is a  primary goal  for the
next-generation facilities and surveys.

As the ionizing  radiation is likely emitted by  hot, young stars, the
current   star-formation    rate   (SFR)   is    of   great   interest
\citep[e.g.][]{bou07}.   However,   the  conversion  from  ultraviolet
luminosity  to  SFR  is   complicated  by  an  unknown  an  extinction
corrections  \citep[e.g.][]{bou10},  which can  be  mitigated to  some
extent with longer wavelength  data \citep{fink10}.  Although with the
Infrared  Array Camera (IRAC)  on the  Spitzer Space  Telescope (\sst)
observations  redward  of  the  4000~\AA-break (in  the  restframe  of
high-redshift  galaxies)   are  routinely  available,   new  practical
problems  with  source  blending  and confusion  have  arisen.   After
dealing  with  issues,  it  seems  that  high-redshift  galaxies  have
$(H-[3.6])\!\sim\!0.6$~mag \citep[e.g.][]{gonz12}.  Na\"{\i}vely, this
suggests that  the galaxies have strong  4000~\AA-breaks indicative of
an  evolved  population \citep[e.g.][]{eyl07},  but  such breaks  seem
unlikely   given  the  age   of  the   Universe  at   these  redshifts
\citep{rich11}.   Instead this red  color may  point to  a significant
emission-line flux in the  IRAC channels \citep{zack08}.  Because both
a 4000~\AA-break  and optical emission  lines are likely  present, the
IRAC photometry  is a critical  component in modeling the  spectra and
determining the  stellar mass, age, and SFR  of high-redshift galaxies
\citep[e.g.][]{pap02}.

Many   of  the   previous  interpretations   of  the   IRAC   data  of
$z\!\gtrsim\!7$ galaxies come from  {\it stacking} fluxes of otherwise
undetected,                     individual                    galaxies
\citep[e.g.][]{labbe10a,gonz12,labbe12}.    In   such  analyses,   one
selects objects of comparable properties (such as $H$-band magnitude),
and  combines  the IRAC  data  to  build  up the  ``average''  signal,
effectively simulating deeper data.  Despite the merits, this approach
has three  short-comings: First, extreme or exotic  objects, which may
challenge existing models or  skew averages, may be excluded.  Second,
this method  implicitly assumes that one obtains  a homogeneous sample
of galaxies by selecting on the $H$-band flux.  However in the case of
$H$-band flux, this is  not guaranteed since this restframe wavelength
is sensitive to both present star formation and extinction.  Therefore
these stacked samples are essentially selected on a combination of SFR
and  dust extinction,  which complicates  the interpretation  of their
``average'' stellar  populations.  Finally, narrow  emission lines can
be  smeared  out  by  stacking  galaxies  of  unknown  (or  imprecise)
redshifts, which complicates the  assessment of their ionizing budget.
In contrast to  stacking, one can use massive  clusters of galaxies as
{\it cosmic telescopes} and magnify background objects, which makes it
possible  to study  intrinsically fainter  individual objects  for the
same exposure  time.  Indeed this  approach is quickly becoming  a key
tool in the study of high-redshift galaxies with the implementation of
the   Hubble   Space   Telescope   (\hst)  Frontier   Fields   program
(HFF)\footnote{http://www.stsci.edu/hst/campaigns/frontier-fields/HDFI\_SWGReport2012.pdf}.

In  this paper, we  present the  first results  from the  {\it Spitzer
  UltRaFaint  SUrvey: \surfsup}, a  {\it Spitzer}  Exploratory Program
(PropID:~90009; PI:~M.~Brada\v{c}) approved in Cycle 9 during the Warm
Mission  \citep{bradac14}.  This  program adds  $\sim\!25$~hr  in both
IRAC channels to  the existing $\sim\!5$~hr for \nclust~strong-lensing
galaxy clusters  at $0.3\!\leq\!z\!\leq\!0.7$.  Six  of these clusters
are  part of  the Cluster  Lensing  and Supernova  Survey with  Hubble
program \citep[CLASH;][]{post12}, two are  scheduled for Year~2 of the
HFF (MACS J0717.5+3745 and MACS J1149.5+2223), and six are planned for
the Grism Lens-Amplified Survey from Space (GLASS; PI:~Treu).  Here we
discuss $z'$-band dropouts lensed  by the Bullet Cluster and identified
by   \citet{hall12}.   This   paper  is   organized  as   follows:  in
section~\ref{sec:obs}    we   discuss    the   \sst/IRAC    data,   in
section~\ref{sec:phot}  we describe  our photometry  and  treatment of
deblending, in section~\ref{sec:sed} we  present the SED modeling, and
in section~\ref{sec:disc} we give concluding remarks with comments for
future work.   We quote all  magnitudes in the  AB system and  adopt a
$\Lambda$CDM      concordance      cosmology      ($\Omega_0\!=\!0.3$,
$\Omega_{\Lambda}\!=\!0.7$, and $H_0\!=\!70$~km~s$^{-1}$~Mpc$^{-1}$).

\section{Observations} \label{sec:obs}

The \sst/IRAC observations  for the Bullet Cluster were  taken as part
of   three   programs  (proposal   IDs:~3550,   60034,  90009),   with
$\gtrsim\!70$\% coming from {\it \surfsup}.  A thorough description of
the   data   reduction   and   survey   strategy   is   discussed   by
\citet{bradac14},  but here  we give  important details.   We generate
mosaics using the \mopex\ software from the corrected-basic calibrated
data   (cBCD)   after    applying   additional   mitigation   measures
\citep[see][]{bradac14}.   There   are  \nframes\  frames,   which  we
drizzled  to an  output  scale of  $0\farcs60$~pix$^{-1}$ yielding  an
integration time  of $\gtrsim\!110$~ks per  pixel in the  regions near
the  \citet{hall12} $z'$-dropouts.   We  astrometrically matched  these
mosaics to  the F160W  images from Wide-Field  Camera~3 (WFC3)  on the
{\it Hubble Space Telescope} (\hst).

In addition  to these IRAC and  existing HST data,  the Bullet Cluster
was also observed by the  {\it Very Large Telescope} (VLT) for 3.75~hr
with the HAWK-I imager  in the $K_s$-band \citep{clement12}.  Although
these  data   have  good  seeing  ($\sim\!0\farcs45$),   none  of  the
\citet{hall12} candidates  are detected.  Instead  we derive $1\sigma$
upper limits from the recovery rate of artificial point-sources placed 
near the positions of the \citet{hall12} sources.

\section{Photometry} \label{sec:phot}

Given  the  dense  cluster  environment,  many  of  the  sources  from
\citet{hall12} are  blended with neighboring  objects, which precludes
the   use   of   simple   aperture  photometry,   therefore   we   use
\pygfit\ \citep{man13}.  We start  by using \galfit\ \citep{peng02} to
model  all objects (with  \sersic\ or  point-source profiles)  for all
objects detected  in \hst/WFC3 F160W  data near the  dropouts.  Taking
these models as  input templates, we extract photometry  from the IRAC
3.6~$\mu$m and 4.5~$\mu$m data.  These $H$-band estimates are taken as
the  initial  conditions  for  \pygfit,  which  convolves  parametric
templates with  a point-spread function (PSF)  and simultaneously fits
for fluxes of all the sources  in the region of interest. We adopt the
empirical PSFs  presented by \citet{bradac14} generated by  a stack of
$\sim\!100$~point sources in the field.   We allow small shifts in the
source coordinates  to account  for any residual  astrometric offsets.
As  a final  note,  the photometry  presented  by \citet{hall12}  were
extrapolated to infinity assuming a model PSF.

\begin{table*}
\caption{Observed Properties}
\label{tab:obs}
\begin{tabular*}{0.98\textwidth}
  {@{\extracolsep{\fill}}ccccccc}
\hline\hline
\multicolumn{1}{c}{galaxy} & \multicolumn{1}{c}{RA} & \multicolumn{1}{c}{Dec} & \multicolumn{1}{c}{$\mu$} & \multicolumn{1}{c}{F160W$^\dagger$} & \multicolumn{1}{c}{[3.6]$^\dagger$} & \multicolumn{1}{c}{[4.5]$^\dagger$}\\
\multicolumn{1}{c}{$ $} & \multicolumn{1}{c}{($^{\mathrm{h}}\;^{\mathrm{m}}\;^{\mathrm{s}}$)} & \multicolumn{1}{c}{($^\circ\;'\;''$)} & \multicolumn{1}{c}{$ $} & \multicolumn{1}{c}{(mag)} & \multicolumn{1}{c}{(mag)} & \multicolumn{1}{c}{(mag)}\\
 & & & & & & \\
\hline
 1 & $06~58~37.13$ & $-55~58~28.070$ & $4.3\pm0.20$ & $26.77\pm0.23$ & $>\!26.7$ & $>\!26.3$ \\
 2 & $06~58~37.26$ & $-55~58~18.844$ & $6.5\pm0.50$ & $26.97\pm0.23$ & $>\!26.5$ & $>\!26.2$ \\
 3 & $06~58~40.17$ & $-55~58~05.041$ & $12\pm4.00$ & $24.97\pm0.16$ & $23.80\pm 0.28$ & $23.78\pm 0.25$ \\
 4 & $06~58~39.30$ & $-55~55~43.687$ & $2.8\pm0.08$ & $26.37\pm0.22$ & $>\!26.5$ & $>\!26.1$ \\
 5 & $06~58~32.25$ & $-55~58~42.971$ & $2.1\pm0.03$ & $25.91\pm0.22$ &  \nodata  &  \nodata  \\
 6 & $06~58~29.87$ & $-55~57~03.834$ & $10.\pm2.00$ & $25.85\pm0.19$ & $>\!26.4$ & $>\!26.1$ \\
 7 & $06~58~34.33$ & $-55~57~53.122$ & $5.2\pm0.50$ & $25.81\pm0.26$ & $>\!26.6$ & $>\!26.2$ \\
 8 & $06~58~34.92$ & $-55~55~29.381$ & $3.1\pm0.10$ & $25.89\pm0.21$ & $>\!26.6$ & $>\!26.2$ \\
 9 & $06~58~31.81$ & $-55~57~49.550$ & $4.3\pm0.30$ & $26.00\pm0.16$ & $>\!26.6$ & $>\!26.2$ \\
10 & $06~58~31.24$ & $-55~58~13.735$ & $3.0\pm0.20$ & $26.37\pm0.16$ & $>\!25.8$ & $>\!24.5$ \\

 & & & & & & \\
\hline
\multicolumn{7}{l}{--The upper limits for [3.6] and [4.5] are 1$\sigma$.}\\
\multicolumn{7}{l}{$^\dagger$These quantities have {\it not} been corrected for the magnification ($\mu$).}
\end{tabular*}
\end{table*}

\begin{figure}
\epsscale{1.2}
\plotone{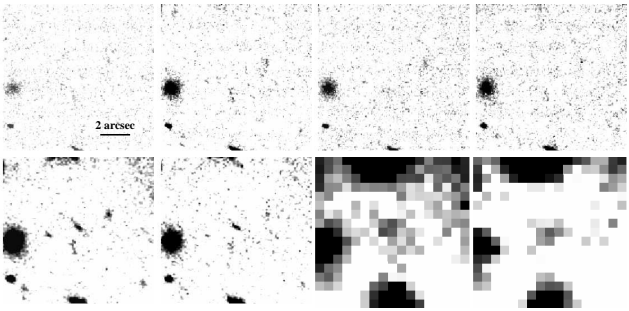}
\caption{\hst\ and \sst\  imaging.  Here we show the  eight images for
  candidate 3 from  \citet{hall12}.  Each stamp is 10''  on a side and
  north-up and  east-left.  The top row shows  the non-detections from
  \hst\ in F606W, F775W,  F814W, and F850LP, respectively.  The bottom
  row shows the dections from  \hst\ and \sst\ in F110W, F160W, [3.6],
  and [4.5],  respectively.  This object is robustly  detected in both
  IRAC   bands   and   reasonably   isolated  from   any   neighboring
  object.  \label{fig:cand3}}
\end{figure}

To estimate the flux uncertainties  from the \galfit\ modeling, we run
simulations in which  we randomly insert point sources  into the image
(all dropouts are  unresolved in IRAC) and compute  the scatter in the
output fluxes as  a function of source brightness.   From the modeling
and simulations, we  derive best fit fluxes and  uncertainties for all
sources  from \citet{hall12} except  for source~5,  which lies  at the
edge of the \hst\  field of view and in IRAC is  blended with a source
outside the  \hst\ footprint.  We  omit this galaxy in  all subsequent
analyses.  In \tab{tab:obs}, we  present our IRAC photometry.  None of
the dropouts  are sufficiently  close in the  IRAC images  to brighter
sources  that confusion  precludes recovery  of the  photometry.  From
this image modeling, we find that only candidate~3 from \citet{hall12}
is robustly  detected ---  it is detected  in both  warm-mission bands
(see    \fig{fig:cand3}).     Whereas    candidate~10,    which    was
spectroscopically confirmed  to be at  $z\!=\!6.740$ \citep{bradac12},
is  not  detected in  either  IRAC  band.   For the  eight  undetected
objects, we estimate upper limits for the IRAC fluxes by computing the
RMS in  a $3\arcsec$ (radius)  aperture on the sky  after ``cleaning''
the  foreground objects.   Our  $1\sigma$ upper  limits are  typically
$[3.6]\!\leq\!26.4$~mag      and      $[4.5]\!\leq\!26.0$~mag     (see
\tab{tab:obs}),  and are not  highly sensitive  to the  aperture size.
These  limits   are  consistent  with   the  exposure-time  calculator
estimates \citet{bradac14} and  only account for the sky  noise in the
vicinity of  the dropouts.   Therefore these limits  are $\sim\!1$~mag
deeper than  the artificial  source tests, which  additionally include
uncertainties  associated with  confusion,  blending, and  overlapping
sources.

\section{SED Modeling and Photometric Redshifts} \label{sec:sed}

\begin{figure}
\epsscale{1.0}
\plotone{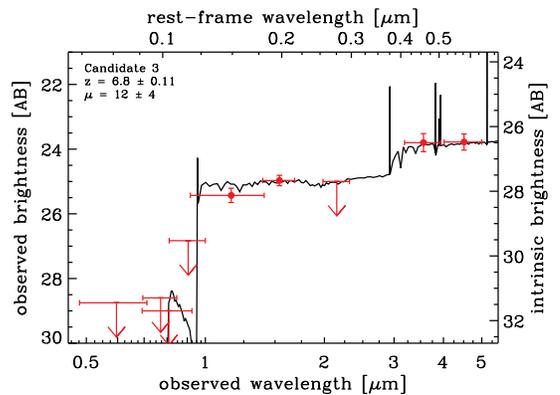}
\caption{SED  fit for  candidate  3.   Here the  red  points show  the
  observed  photometry  from  \hst/VLT/\sst\  (the  upper  limits  are
  $1\sigma$),  and   the  black  line  is  the   best-fit  model  from
  \lephare\ \citep[including  emission lines][]{ilb09}.  On  the right
  vertical  axis,  we show  the  {\it  intrinsic} apparent  magnitudes
  ($m_{\rm  inst}\!=\!m_{\rm   obs}+2.5\log\mu$)  to  demonstrate  the
  effect of the lensing.  The horizontal error bars represent the FWHM
  and the points are placed at  the mean wavelength of each band.  For
  this    object,    we    obtain    a   photometric    redshift    of
  $z\!=\!6.8_{-0.10}^{+0.12}$  which   gives  an  FUV   luminosity  of
  $L_{1500}\!\sim\!0.3~L^*_{z\!=\!7}$, the lowest-luminosity source at
  $z\!\gtrsim\!6$ yet detected by  \sst/IRAC.  Constraining the SFR of
  such  low-luminosity  galaxies   is  critical  in  establishing  the
  ionizing flux budget.\label{fig:sed}}
\end{figure}

For  each candidate,  we fit  the combined  nine-band  photometry from
\hst, VLT,  and \sst\ using \lephare\ \citep{ilb06,ilb09}.   We use 27
stellar population synthesis models generated using \citet[][hereafter
  BC03]{bc03}  models at 58  ages that  range from  $0.1-13.5$~Gyr and
exclude  ages  greater  than   the  age  of  the  Universe\footnote{At
  $z\!\sim\!7$ the  age of  the Universe is  $\sim\!760$~Myr.}.  These
BC03  models use  a \citet{chab03}  initial mass  function  (IMF) with
metallicities of  $Z\!=\!(0.02-1)~Z_{\odot}$ and are  characterized by
exponentially-declining    star-formation   histories    (SFHs)   with
timescales  of  $\tau\!=\!0.1-30$~Gyr.   We adopt  the  \citet{calz00}
reddening   law  with   $0\!\leq\!E(B-V)\!\leq\!0.5$.    In  addition,
\lephare\ adds  nebular emission lines  to the BC03 templates  using a
direct  translation   between  the  dust-corrected   ultraviolet  (UV)
luminosity    at   rest-frame    2300~\AA\   and    the   [\ion{O}{2}]
$\lambda3727$~\AA\ feature,  which is subsequently used  to derive the
strength  of  other  rest-frame  UV/optical  emission  features  using
average  intrinsic ratios  \citep[see][]{ilb09}.   \lephare\ estimates
the  current  SFR  from  dust-corrected  UV  luminosity  and  standard
relations \citep{kenn98}. To  estimate parameter uncertainties, we add
Gaussian noise to each photometric  point for each galaxy, re-fit with
\lephare,   and  compute   the   RMS  of   best-fit  parameters.    In
\tab{tab:pop},  we  present  the   best-fit  SED  parameters  for  the
\citet{hall12}   $z'$-dropouts  and   show   the  SED   fit  for   the
IRAC-detected object in \fig{fig:sed}.

\begin{table}
\caption{Stellar Population Results}
\label{tab:pop}
\begin{tabular*}{0.48\textwidth}
  {@{\extracolsep{\fill}}cccccc}

\hline\hline
\multicolumn{1}{c}{gal} & \multicolumn{1}{c}{$z$} & \multicolumn{1}{c}{$\log~t$} & \multicolumn{1}{c}{$\log~M^\dagger$} & \multicolumn{1}{c}{$\log$~SFR$^\dagger$} & \multicolumn{1}{c}{$E_{B-V}$}\\
\multicolumn{1}{c}{$ $} & \multicolumn{1}{c}{$ $} & \multicolumn{1}{c}{(yr)} & \multicolumn{1}{c}{(M$_\odot$)} & \multicolumn{1}{c}{(M$_\odot$~yr$^{-1}$)} & \multicolumn{1}{c}{(mag)}\\
 & & & & & \\
\hline

$1$ & $6.5_{-0.20}^{+0.30}$ & $7.6_{-0.58}^{+0.55}$ & $7.9_{-0.28}^{+0.20}$ & $0.24_{-0.34}^{+0.34}$ & $0.10_{-0.1}^{+0.1}$ \\
$2$ & $6.4_{-0.27}^{+0.17}$ & $7.7_{-0.68}^{+0.81}$ & $7.7_{-0.62}^{+0.33}$ & $-0.10_{-0.34}^{+0.62}$ & $0.0_{-0.0}^{+0.2}$ \\
$3$ & $6.8_{-0.10}^{+0.12}$ & $8.8_{-0.20}^{+0.10}$ & $9.3_{-0.23}^{+0.11}$ & $0.12_{-0.28}^{+0.31}$ & $0.0_{-0.0}^{+0.4}$ \\
$4$ & $7.9_{-0.87}^{+0.065}$ & $7.5_{-0.46}^{+0.38}$ & $8.1_{-0.23}^{+0.19}$ & $0.56_{-0.21}^{+0.41}$ & $0.0_{-0.0}^{+0.1}$ \\
$5$ & \multicolumn{5}{c}{omitted from this analysis} \\
$6$ & $7.1_{-0.10}^{+0.75}$ & $7.0_{-0.00}^{+0.72}$ & $7.3_{-0.06}^{+0.065}$ & $0.22_{-0.08}^{+0.054}$ & $0.0_{-0.0}^{+0.1}$ \\
$7$ & $7.0_{-0.10}^{+1.0}$ & $7.0_{-0.00}^{+0.89}$ & $7.4_{-0.06}^{+0.041}$ & $0.43_{-0.08}^{+0.041}$ & $0.0_{-0.0}^{+0.1}$ \\
$8$ & $6.9_{-0.40}^{+1.1}$ & $7.0_{-0.00}^{+1.0}$ & $7.7_{-0.08}^{+0.16}$ & $0.61_{-0.12}^{+0.077}$ & $0.0_{-0.0}^{+0.1}$ \\
$9$ & $7.8_{-0.58}^{+0.17}$ & $7.0_{-0.00}^{+0.38}$ & $7.6_{-0.06}^{+0.13}$ & $0.52_{-0.12}^{+0.076}$ & $0.0_{-0.0}^{+0.1}$ \\
$10\tablenotemark{*}$ & $6.740\tablenotemark{*}$ & $7.7_{-0.63}^{+0.65}$ & $8.1_{-0.51}^{+0.55}$ & $0.43_{-0.19}^{+0.26}$ & $0.0_{-0.0}^{+0.3}$ \\
 & & & & & \\
\hline
\multicolumn{6}{l}{\hspace*{4pt}The SFRs presented here was derived with \lephare, and is}\\
\multicolumn{6}{l}{\hspace*{10pt}similar to the calculation discussed in \S~\ref{sec:disc}.}\\
\multicolumn{6}{l}{$^*$For this object, the redshift was fixed to the spectroscopic value}\\
\multicolumn{6}{l}{\hspace*{6pt}\citep{bradac12}.}\\
\multicolumn{6}{l}{$^\dagger$These quantities have been corrected for the magnification ($\mu$,}\\
\multicolumn{6}{l}{\hspace*{6pt}see \tab{tab:obs}).}\\
\end{tabular*}
\end{table}

\begin{figure}
\epsscale{1.2}
\plotone{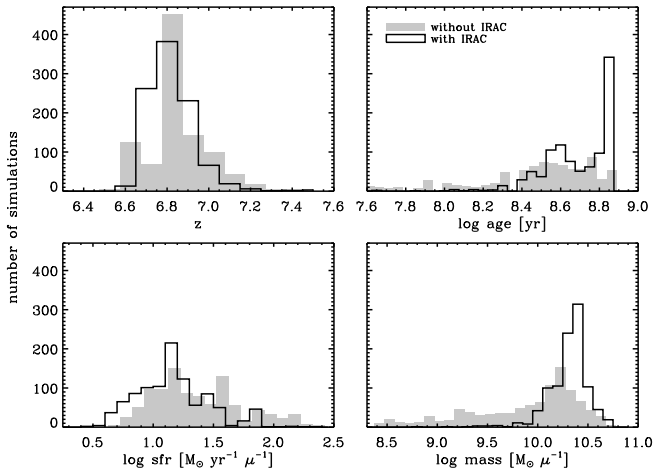}
\caption{Stellar population  parameters for candidate  3.  To estimate
  the uncertainty  on these  parameters, we use  a simple  Monte Carlo
  simulation  where we  fit each  galaxy  1000 times  with the  fluxes
  tweaked by  their uncertainties.   The open histogram  shows results
  for the nine-band photometry  from \hst/VLT/\sst.  To illustrate the
  importance of the IRAC data  in modeling these galaxies, we show the
  results without the IRAC  data (shaded histogram, we slightly offset
  the shaded histograms  for clarity).  The redshift is  robust to the
  exclusion  of the  IRAC  data, whereas  the  SFR is  $\sim\!2$~times
  higher  without   IRAC  and   stellar  mass  is   almost  completely
  uncertain.\label{fig:spp}}
\end{figure}

In \fig{fig:spp}  we show the  distribution of SED parameters  for the
object~3,  which was  detected in  both IRAC  channels.  The  open and
shaded histograms show  the results with and without  the inclusion of
the  IRAC  data,  respectively.   The  rest-frame  optical  data  help
markedly in constraining  the age and stellar mass,  whereas they have
less effect on the redshift or the precision of the SFR rate.  This is
not   surprising  because   much  of   the  redshift   information  is
encapsulated  in  the  observed  Lyman  break between  the  $z'$-  and
$J$-bands, whereas the  age and stellar mass depend  critically on the
4000~\AA\  break and  the  rest-frame optical  data.   Finally, it  is
becoming  widely  accepted  that  optical  emission  lines  (H$\beta$,
[\ion{O}{3}],  H$\alpha$) contribute  a significant  flux to  the IRAC
bands.   However, object~3  seems to  be a  unique redshift  where the
additional flux from H$\beta$ and [\ion{O}{3}] is offset by a slightly
lower continuum, giving a fairly  ``neutral'' IRAC color.  Using a set
of              fixed             equivalent             widths:~${\rm
  EW_{rest}}$(H$\beta$,[\ion{O}{3}],H$\alpha)\!=\!(105,670,300)$~\AA,
\citet{gonz12}  show  that emission  lines  contribute  little to  the
observed  $([3.6]-[4.5])$ color.  Therefore  the equivalent  widths of
these  emission  lines in  object~3  are  likely  no larger  than  the
\citet{gonz12} assumptions.

\section{Discussion} \label{sec:disc}

We      have      presented      the      first      results      from
\surfsup\footnote{http://www.physics.ucdavis.edu/$\sim$marusa/SurfsUp.html},
a  Spitzer  Exploration   Program  to  image  \nclust\  strong-lensing
clusters to  $\sim\!100$~ks depth  per channel.  We  have definitively
detected   one  of   the  \ngal\   $z'$-band  dropouts   identified  by
\citet{hall12}.  This galaxy is highly magnified by the Bullet Cluster
($\mu\!=\!12\pm4$)      with     an     apparent      magnitude     of
$J\!=\!25.43\pm0.22$~mag,   which  gives   a   far-UV  luminosity   of
$M_{1500}\!=\!-18.9\pm0.42$~mag  (accounting for both  the photometric
and   magnification   uncertainty).    Therefore   this   galaxy   has
$L_{1500}\!\sim\!0.3\;L^*$  \citep[taking  $M_{1500}^*\!=\!-20.14$~mag
  from][]{bou11},  and is the  only {\it  individual} dwarf  galaxy at
$z\!\gtrsim\!7$ detected so far by IRAC \citep[c.f.][]{yan12,labbe12}.
This  is the  first  direct detection  of  the kind  of galaxy  likely
responsible  for the cosmic  reionization.  From  the SED  modeling we
infer    a   specific    star-formation   rate    (sSFR)    of   ${\rm
  sSFR}\!\sim\!0.7$~Gyr$^{-1}$,   which  is   lower   than  comparable
galaxies    \citep[e.g.][]{zheng12,zit12}   or    at    low   redshift
\citep[e.g.][]{kai}.  In  contrast, the remaining  eight galaxies have
$\langle{\rm     sSFR}\rangle\!\sim\!50$~Gyr$^{-1}$,     similar    to
Lyman-break galaxies (LBGs)  at $z\!\sim\!5$ \citep{nph12}, suggesting
that  this   detected  object  may   be  an  unusual  member   of  the
high-redshift galaxy population.

It is intriguing  to consider the SFHs that  could yield a substantial
stellar  mass  ($M_*\!\sim\!2^{+0.6}_{-0.9}\times10^9$~M$_{\odot}$) at
this  early  epoch.   The  two  inferences of  stellar  mass  and  SFR
essentially  constrain the  integral  and current  value  of the  SFH,
respectively.   Assuming that  galaxies begin  to form  around $z_{\rm
  form}\!\sim\!20$, then  this galaxy must have  acquired the observed
stellar mass in  $\lesssim\!650$~Myr\footnote{Although we estimate the
  age   in    \tab{tab:pop},   this   age   is    predicated   on   an
  exponentially-declining SFH.  To  avoid circularity in the argument,
  we  instead   adopt  a  conservative  estimate   for  the  formation
  redshift.}.  If it had constantly  formed stars at the measured rate
over  this  time, then  it  would  have built  up  a  stellar mass  of
$8.4\times10^8$~M$_{\odot}$.   Although  this  constant SFR  model  is
roughly  consistent  with the  derived  mass ($\sim\!1.5~\sigma$),  it
suggests that  the SFR could not  have been lower in  the past without
some corresponding  period of increased star formation.   Of course it
is  impossible to  distinguish between  a smooth,  multi-component SFH
\citep[e.g.][]{lee10,behroozi,pacifici}  from   a  stochastic  history
punctuated  by   intense  bursts.   But,   it  does  imply   that  the
exponentially-increasing model \citep{mara10}  can be ruled out, given
its substantial stellar  mass, modest SFR, and high  redshift.  If the
actual formation redshift were lower than our conservative assumption,
then  the argument  becomes  stronger as  the constant  star-formation
scenario cannot create enough mass by $z\!\sim\!7$.

In the above we tacitly assumed that the stellar mass was created {\it
  in  situ}, and  that  it had  not  experienced any  type of  merger.
Although mergers would bring in stellar mass (and possibly enhance the
star formation),  they are not on  frequent enough to  change the mass
significantly \citep{hopk10}.  Using  their merger rate calculator, we
estimate    that    a    galaxy     with    a    stellar    mass    of
$2\times10^9$~M$_{\odot}$ will  have an  average major merger  rate of
$\sim\!0.9$~mergers~Gyr$^{-1}$.   In  the  $600-700$~Myr available  to
this galaxy  (the range reflects  the $\pm3\sigma$ uncertainty  on the
photometric redshift), there are  only $0.5-0.7$~mergers of mass ratio
$0.25\!<\!m_1/m_2\!<1$.   Even  if major  mergers  contributed to  the
observed  stellar  mass  of  this  galaxy, this  scenario  raises  the
question   of   how   two   progenitors   with   stellar   masses   of
$M_*\!\sim\!10^9$~$M_{\odot}$  formed --- the  one galaxy  is puzzling
enough.  While this  galaxy may have had an  exceptionally high merger
rate,  the major merger  scenario seems  unlikely, which  leaves minor
mergers  or  gradual accretion/inflow  a  possibility.  Certainly  gas
inflow  is an  early  prediction  for the  formation  of the  earliest
galaxies  \citep[e.g.][]{larson}   and  has  even   been  observed  in
low-redshift  star-forming  galaxies  \citep{rubin12}.  Our  broadband
data  are  insensitive  to  the  observational  signatures  of  inflow
\citep[such as redshifted resonance lines;][]{martin}, but it possible
with the {\it The James Webb Space Telescope}.

%the ALMA stuff....
We      estimate      this       galaxy      has      an      observed
$\mbox{SFR}\!=\!\psilensed\,\mu^{-1}$~M$_{\odot}$~yr$^{-1}$,  and is a
prime candidate for follow-up  with the Atacama Large Millimeter Array
(ALMA).  Like  the far-ultraviolet  (FUV), the far-infrared  (FIR) has
many   useful   SFR    indicators,   particularly   the   [\ion{C}{2}]
$\lambda157.7~\mu$m  emission  line and  thermal  continuum from  warm
dust.  To predict  the [\ion{C}{2}] flux, we use  the SFR estimated in
Section~\ref{sec:sed}  and  calibrations  from  \citet{delooze11},  to
predict    an   integrated    line   flux    of    $F_{[\rm   C~{\tiny
      II}]}\!\sim\!\ciiflux$.   If we assume  a Gaussian  line profile
(of width  $\Delta v\!=\!100$~km~s$^{-1}$),  the SFR gives  an average
flux       density      of       $\langle       f_{[{\rm      C~{\tiny
        II}}]}\rangle\!=\!\aveciiflux$.  Because  our SED models imply
a  small  amount  of  dust  (see \tab{tab:pop})  and  recent  chemical
evolution models find  a sufficient amount of dust  can be produced by
$z\!\sim\!6$  \citep{vali09},  we  estimate  the  FIR  continuum  flux
(8--1000~$\mu$m) using the the \citet{kenn98} scaling relation.
%of     $L_{\rm    FIR}~({\rm    erg~s}^{-1})\!=\!2.2\times10^{43}~{\rm
%  SFR}~(M_{\odot}~{\rm yr}^{-1})$.
We  predict  this galaxy  will  have  an  integrated flux  of  $F_{\rm
  FIR}\!=\!\firflux$,   which    averaged   over   8--1000~$\mu$m   is
$\left<f_{\rm  FIR}\right>\!=\!\avefirflux$.  All fluxes  discussed in
this  paragraph  include the  magnification.   These  flux levels  are
readily   achievable,   even   with   the  current   ALMA   facilities
\citep[e.g.][]{wagg12}.

Using  standard tools  and techniques  we have  robustly  detected 1/9
$z'$-dropouts from  \citet{hall12}.  For the  remaining eight galaxies
that  have  gone  undetected  in  \surfsup\  we  performed  a  similar
tweaking/re-fitting  analysis described in  section~\ref{sec:sed}.  We
tweak the IRAC upper limits for the eight undetected galaxies slightly
deeper  ($\sim\!0.25$~mag) and  re-fit  with \lephare.   We find  that
either  no  combination   of  physical  parameters  could  effectively
characterize the tweaked photometry  or that resulting best-fit models
required extreme  properties: very young  ages ($\lesssim\!10$~Myr) or
low  stellar  masses  ($\lesssim\!10^8$~M$_{\odot}$~$\mu^{-1}$;  where
$\mu$ is the magnification).  As such, we suspect that the IRAC limits
quoted in \tab{tab:obs} for the remaining eight galaxies (where one is
unobservable  as  discussed in  section~\ref{sec:phot})  are close  to
their  true  brightnesses.   As  part  of our  on-going  efforts  with
\surfsup\ we  will continue  to develop tools  and techniques  to deal
with the unique challenges posed by this and similar datasets.

\acknowledgments  We would  like to  thank the  anonymous  Referee for
several  insightful  suggestions  and  comments.   We  also  recognize
Olivier  Ilbert, Steve  Finkelstein,  Seth Cohen,  Norman Grogin,  Jen
Lotz,  and Preethi  Nair for  their helpful  discussions.   The HAWK-I
$K$-band data was provided by B.~Clement, and was taken as part of the
ESO Large Program ID:~181.A-0485 (PI:~J.~G.~Cuby).  This work is based
[in part] on observations made with the Spitzer Space Telescope, which
is operated by the  Jet Propulsion Laboratory, California Institute of
Technology  under a  contract with  NASA.  Support  for this  work was
provided by  NASA through an  award issued by JPL/Caltech  and through
HST-GO-10200,   HST-GO-10863,  and   HST-GO-11099   from  STScI.    TT
acknowledges support by the Packard  Fellowship.  Part of the work was
carried out by MB and  TT while attending the program ``First Galaxies
and Faint Dwarfs'' at KITP which is supported in part by the NSF under
Grant No.~NSF  PHY11-25915.  TS  acknowledges support from  the German
Federal Ministry  of Economics and Technology  (BMWi) provided through
DLR under  project 50 OR  1308.  HH is  supported by the DFG  grant Hi
1495/2-1.   SA and  AvdL  acknowledge support  by  the U.S.~DoE  under
contract  number DE-AC02-76SF00515  and by  the Dark  Cosmology Centre
which is funded by the Danish National Research Foundation.

{\it Facilities:} \facility{SST (IRAC)}

\end{document}